# The impact of a nation-wide lockdown on COVID-19 transmissibility in Italy

Giorgio Guzzetta[a,*], Flavia Riccardo[b,*], Valentina Marziano[a], Piero Poletti[a], Filippo Trentini[a], Antonino Bella[b], Xanthi Andrianou[b,c] Martina Del Manso[b], Massimo Fabiani[b], Stefania Bellino[b], Stefano Boros[b], Alberto Mateo Urdiales[b], Maria Fenicia Vescio[b], Andrea Piccioli[b], COVID-19 working group[§], Silvio Brusaferro[b], Giovanni Rezza[b], Patrizio Pezzotti[b,#], Marco Ajelli[a,#,^], Stefano Merler, MSc[a,#]

[a] : Fondazione Bruno Kessler, Trento, Italy
[b] : Istituto Superiore di Sanità, Rome, Italy
[c] : Cyprus University of Technology, Limassol, Cyprus

[*] : equal contributions
[#] : joint senior authors
[^] : corresponding author. Dr. Marco Ajelli, Bruno Kessler Foundation, 38123 Trento, Italy. E-mail: marco.ajelli@gmail.com

[§] : Members of the COVID-19 working group Maria Rita Castrucci, Alessandra Ciervo, Fortunato (Paolo) D'Ancona, Corrado Di Benedetto, Antonietta Filia, Stefania Giannitelli, Ornella Punzo, Maria Cristina Rota, Andrea Siddu, Paola Stefanelli, Marco Tallon, Roberta Urciuoli (Istituto Superiore di Sanità).Regional representatives: Antonia Petrucci (Abruzzo); Michele Labianca(Basilicata); Anna Domenica Mignuoli (Calabria); Angelo D'Argenzio (Campania); Erika Massimiliani (Emilia-Romagna); Tolinda Gallo (Friuli Venezia Giulia); Paola Scognamiglio (Lazio); Camilla Sticchi (Liguria); Danilo Cereda (Lombardia); Daniel Fiacchini (Marche); Francesco Sforza (Molise); Maria Grazia Zuccaro (P.A. Bolzano); Pier Paolo Benetollo (P.A. Trento); Donatella Tiberti (Piemonte); Maria Chironna (Puglia); Maria Antonietta Palmas (Sardegna); Salvatore Scondotto (Sicilia); Emanuela Balocchini (Toscana); Anna Tosti (Umbria); Mauro Ruffier (Valle D'Aosta); Filippo Da Re (Veneto).

**Abstract**

On March 10, 2020, Italy imposed a national lockdown to curtail the spread of COVID-19. Here we estimate that, fourteen days after the implementation of the strategy, the net reproduction number to has dropped below the epidemic threshold – estimated range 0.4-0.7. Our findings provide a timeline of the effectiveness of the implemented lockdown, which is relevant for a large number of countries that followed Italy in enforcing similar measures.

On February 21, 2020, the first case of locally transmitted COVID-19 was detected in Italy, in a 37-year-old man residing in Lombardy [1, 2]. Since then, a number of interventions have been deployed to contain disease spread, initially on a geographic basis and eventually culminating in a nationwide lockdown since March 11, 2020 [3, 4]. At the time of writing, the national lockdown is still in force.

**Net reproduction number**

To evaluate the impact of performed interventions, we derive updated estimates of daily disease transmissibility, measured in terms of the net reproduction number R. This quantity represents the mean number of secondary infections generated by one primary infector in the presence of control interventions and human behavioral adaptations. When R decreases below the epidemic threshold of 1, the number of new infections begins to decline. The objective of this study was to evaluate R at the end of March 2020 in Italy.

Case-based surveillance data were collected by regional health authorities and collated by the Istituto Superiore di Sanità using a secure online platform, according to a progressively harmonized track-record. Data include, among other information, the place of residence, the date of symptom onset and the date of first hospital admission for laboratory-confirmed COVID-19 cases [4].

The distribution of the net reproductive number R(t) was estimated by applying a well-established statistical method [5-7], which is based the knowledge of the distribution of the generation time and on the time series of cases. In particular, the posterior distribution of R for any time point t was estimated by applying the Metropolis-Hastings MCMC sampling to a likelihood function defined as follows:

$$\mathcal{L} = \prod_{t=1}^{T} P\left(C(t); R(t) \sum_{s=1}^{T} \varphi(s) C(t-s)\right)$$

where

- $P(k; \lambda)$ is the probability mass function of a Poisson distribution (i.e., the probability of observing k events if these events occur with rate λ).
- C(t), is the daily number of new cases (imported or locally acquired) at time t;
- R(t) is the net reproduction number at time t to be estimated;
- $\varphi(s)$ is the distribution of the generation time calculated at time s.

As a proxy for the distribution of the generation time, we used the empirical distribution of the serial interval, estimated from the analysis of contact tracing data in Lombardy [4], i.e., a gamma function with shape 1.87 and rate 0.28, having a mean of 6.6 days. This estimate is within the range of other available estimates for SARS-CoV-2 infections, i.e. between 4 and 7.5 days [8-10].

We computed the value of R(t) from two different data sets: (i) the time series of COVID-19 cases by date of symptom onset (estimate denoted by $R^{symp}$) and (ii) the time series of hospitalized cases by date of hospital admission, H(t) (estimate denoted by $R^{hosp}$). As cases are admitted to the hospital at delayed time D from their symptom onset, we computed $R^{hosp}(t)$ using the shifted time series of hospitalized cases, H(t+D). The median value of D was estimated at 7 days from surveillance data, using 32,893 cases for which both the date of symptom onset and the date of hospital admission were available.

To account for the geographical heterogeneity in contacts, healthcare organization, timelines of interventions, and compliance to social distancing and use of precautions for transmission reduction, R was estimated separately for different provinces and regions. We included in the analysis provinces and regions having a proportion of cases with a recorded symptom onset date of at least 40% with respect to the overall number of reported cases.

Overall, the selected provinces and regions covered a total population of 37 million inhabitants, 61.3% of the Italian population and accounted for 61,707 symptomatic cases (90.4% of the total) and 34,708 (87.0% of the total) hospitalizations.

**Effect of the lockdown on the net reproduction number**

We found that the estimated values for $R^{symp}$ at March 25 were consistently below the epidemic threshold of 1 for all the analyzed regions and autonomous provinces (Figure 1). The 95% confidence interval (CI) of the posterior estimate for $R^{hosp}$ exceeded the unit only for Apulia and Sardinia. These two regions reported the lowest relative cumulative incidence among the included ones (27 per 100,000, compared for instance to 322 per 100,000 for Lombardy – the most affected region).

The mean value of R across the regions and autonomous provinces, weighted by the number of reported cases at March 25, was 0.7 for $R^{symp}$ and 0.46 for $R^{hosp}$. Results were consistent when analyzing estimates from the 57 selected provinces, belonging to 14 unique regions. The mean value of R was below the epidemic threshold for 84% of provinces when using the time series of symptom onset dates and for 88% when using hospital admission dates. The upper limit of the confidence interval was below 1 for 58% of the provinces when using $R^{symp}$ and 60% when using $R^{hosp}$. Provinces with mean R above 1 reported a significantly lower number of cases (average: 163 cases and 77 hospitalizations) compared to provinces with mean R below 1 (average: 1,215 cases and 681 hospitalizations; t-test p-values < 0.005). The mean value of R across the provinces, weighted by the province's number of reported cases at March 25, was 0.68 for $R^{symp}$ and 0.40 for $R^{hosp}$. Figure 1 also shows that the value of R estimated with the two methods was consistently and largely above the epidemic threshold at March 10 in all considered regions.

**Discussion**

Our results suggest that the restrictive interventions put in place to limit the spread of SARS-CoV-2 in Italy have been successful in bringing the reproduction number significantly below 1 within two weeks from the national lockdown on March 11, 2020. As of March 25, we estimated a mean reproduction number between 0.6 and 0.8 for the same regions. At March 10, the value of the net reproduction number 2020 ranged between 1.5 and 3.2 in the same regions (Figure 1 and [4]). Notably, the reproductive number had been declining steeply, although insufficiently, even before the national lockdown, especially in the hard-hit regions of Lombardy and Emilia-Romagna, thanks to the initial geographically targeted measures taken after the discovery of the epidemics [4]. The lockdown was fundamental to prevent an explosion in the number of cases in regions that had a great epidemic potential by March 10th, but where transmission had started weeks later compared to the outbreak epicenter.

One limitation of the method for estimating the reproduction number consists in the requirement that the case notification rate remains constant over time. However, during the course of the epidemic, some regions had to face severe limitations in testing capacity and were forced to change the criteria for testing accordingly. If the result of these changes was a reduction of the notification rate over time, the reproduction number might be underestimated [11]. On the other hand, a massive and sustained scale-up of testing capacity was also set up in all regions during the course of the epidemics (Figure 3) [12], which was not accompanied by a corresponding increase of confirmed incident cases in the weeks following March 25. This suggests an increase of notification rates and thus a possible overestimation of R [11]. To compensate possible biases, we supplemented our estimates of the reproduction number using alternative estimates from the time series of hospitalized cases. Criteria for hospitalization are more homogeneous across the local health systems and over time than testing criteria, as they are grounded on the patient's need for medical assistance. Furthermore, the hospitalization date is easier and more reliable information to collect with respect to the symptom onset date, which requires an epidemiological investigation and may be subject to recall bias. Results obtained with this additional method further support our conclusions.

On the same day that the WHO declared SARS-CoV-2 a pandemic [13], Italy was the first country in the western hemisphere to impose a nationwide lockdown, although with softer restrictions compared to the Chinese experience. Many countries worldwide followed Italy in the same decision with a delay of a few days to a few weeks and similar degrees of enforcement. The effectiveness of lockdown interventions had been shown in China, where the reproduction number was estimated to fall to about 0.3 in Wuhan [14] and 0.5 in other Chinese provinces [11]. Here we demonstrate that the comparatively lighter measures implemented by Italy were capable to achieve rapidly control of the epidemics, although at probably higher values of the reproduction number. Whether residual viral circulation will result in new epidemic waves after lockdown removal remains undefined. These results are of high importance for worldwide efforts to contain the first wave of the SARS-CoV-2 pandemic.

**Conflict of interest**

The authors declare no conflict of interest.

**Funding statement**

GG, VM, PPo, FT, MA, and SM acknowledge funding from the European Commission H2020 project MOOD and from the VRT Foundation Trento project "Epidemiologia e transmissione di COVID-19 in Trentino".

**References**

1. Cereda D, Tirani M, Rovida F, Demicheli V, et al. The early phase of the COVID-19 outbreak in Lombardy, Italy. ArXiv200309320 Q-Bio. 20 Mar 2020; Available at: http://arxiv.org/abs/2003.09320
2. Guzzetta G, Poletti P, Ajelli M, Trentini F, et al. Potential short-term outcome of an uncontrolled COVID-19 epidemic in Lombardy, Italy, February to March 2020. Eurosurveillance. 2020;25(12):2000293.
3. Decree of the Prime Minister. Ulteriori disposizioni attuative del decreto-legge 23 febbraio 2020, n. 6, recante misure urgenti in materia di contenimento e gestione dell'emergenza epidemiologica da COVID-19, applicabili sull'intero territorio nazionale. (20A01605) (G.U. Serie Generale , n. 64 del 11 marzo 2020) Internet]. 2020. Available at: http://www.trovanorme.salute.gov.it/norme/dettaglioAtto?id=73643
4. Riccardo F, Ajelli M, Andrianou X, Bella A, et al. Epidemiological characteristics of COVID-19 cases in Italy and estimates of the reproductive numbers one month into the epidemic. Medrxiv preprint https://doi.org/10.1101/2020.04.08.20056861
5. WHO Ebola Response Team. Ebola virus disease in West Africa—the first 9 months of the epidemic and forward projections. *N Engl J Med.* 2014; 371: 1481-1495.
6. Cori A, Ferguson NM, Fraser C, Cauchemez S. A new framework and software to estimate time-varying reproduction numbers during epidemics. American journal of epidemiology. 2013;178(9):1505-12.
7. Liu Q-H, Ajelli M, Aleta A, Merler S, Moreno Y, Vespignani A. Measurability of the epidemic reproduction number in data-driven contact networks. Proc Natl Acad Sci. 2018; 115(50):12680–5.
8. Nishiura H, Linton NM, Akhmetzhanov AR. Serial interval of novel coronavirus (COVID-19) infections. International Journal of Infectious Diseases. 2020.
9. Wu JT, Leung K, Bushman M, Kishore N, Niehus R, de Salazar PM, Cowling BJ, Lipsitch M, Leung GM. Estimating clinical severity of COVID-19 from the transmission dynamics in Wuhan, China. Nature Medicine. 2020.
10. Li Q, Guan X, Wu P, Wang X, et al. Early transmission dynamics in Wuhan, China, of novel coronavirus–infected pneumonia. New England Journal of Medicine. 2020.

11. Zhang J, Litvinova M, Wang W, Wang Y, et al. Evolving epidemiology and transmission dynamics of coronavirus disease 2019 outside Hubei province, China: a descriptive and modelling study. The Lancet Infectious Diseases. 2020.
12. Dipartimento di Protezione Civile. COVID-19 Italia - Monitoraggio situazione. Data repository at https://github.com/pcm-dpc/COVID-19
13. WHO Director-General's opening remarks at the media briefing on COVID-19. https://www.who.int/dg/speeches/detail/who-director-general-s-opening-remarks-at-the-media-briefing-on-covid-19---11-march-2020
14. Pan A, Liu L, Wang C, Guo H, et al. Association of public health interventions with the epidemiology of the COVID-19 outbreak in Wuhan, China. JAMA. 2020.

**Figures**

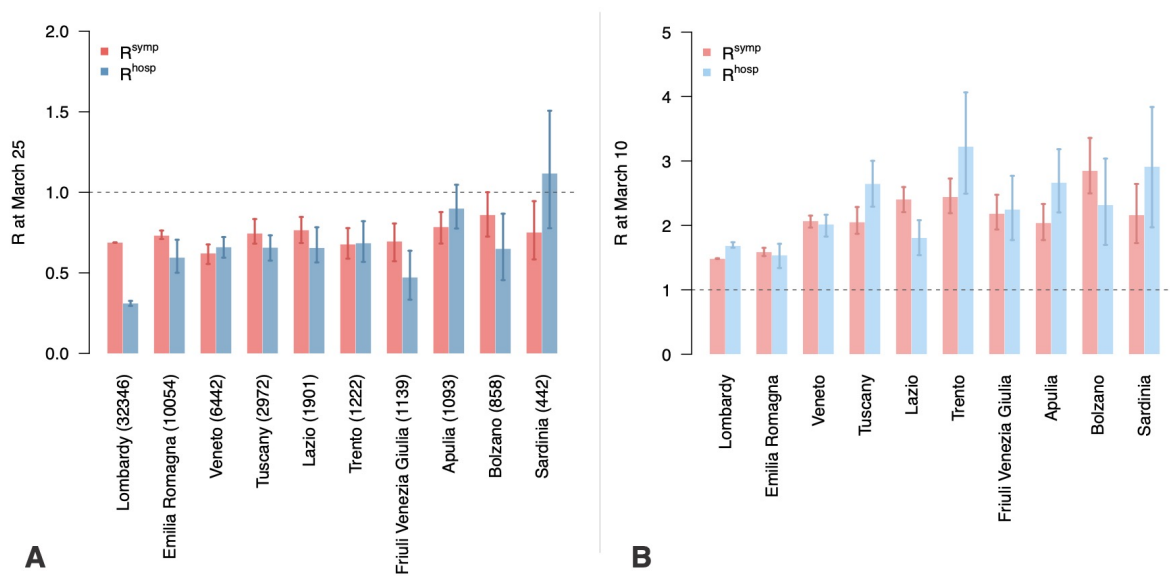

**Figure 1.** Estimates of the reproduction numbers $R^{symp}$ and $R^{hosp}$ for 8 selected Italian regions and the two autonomous provinces of Trento and Bolzano. Regions are sorted by decreasing absolute cumulative incidence at March 25 (reported in parentheses after the region name). **A** Estimates at March 25th. **B** Estimates at March 10th. Solid bars and vertical lines: mean and 95% CI from the posterior distribution. The horizontal dashed grey line represents the epidemic threshold.

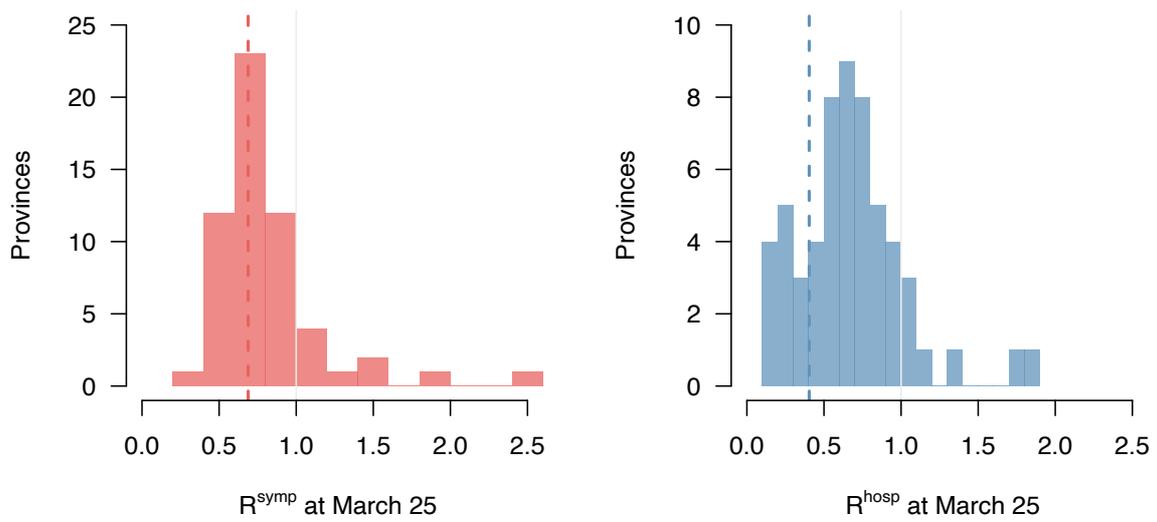

**Figure 2.** Estimates of the reproduction numbers $R^{symp}$ and $R^{hosp}$ at March 25$^{th}$, for 57 selected Italian provinces. Dashed vertical lines: average of R, weighted by the number of reported cases by each province.

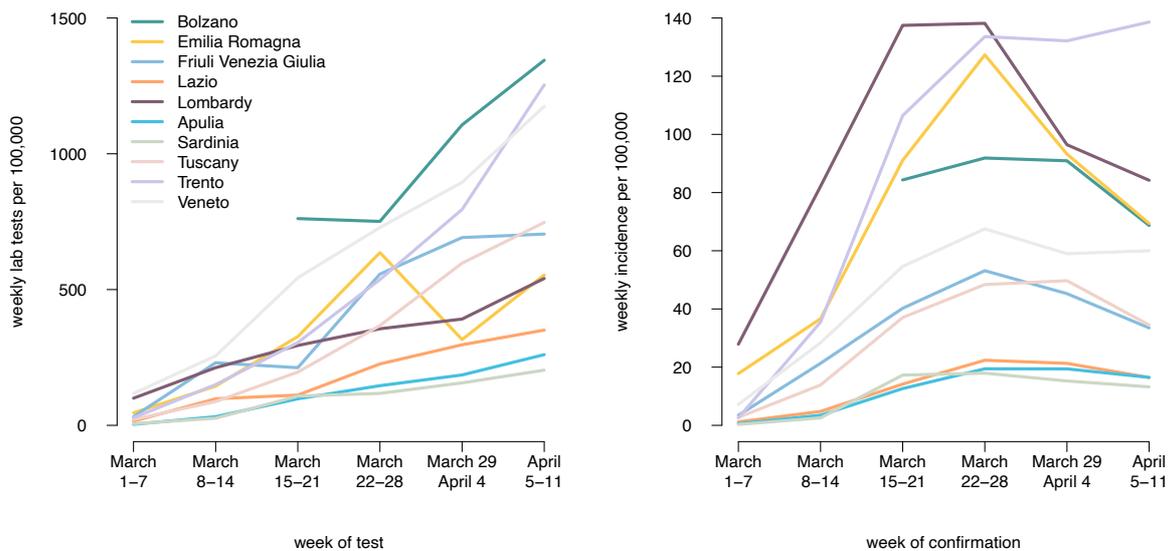

**Figure 3.** Number of lab tests and lab-confirmed incident cases per 100,000 population in the 8 selected Italian regions and the two autonomous provinces, as reported by the Italian Dipartimento di Protezione Civile [12]. Note that confirmed cases refer to infections occurring several days and up to few weeks before, due to delays related to development of symptoms, seeking for medical assistance, execution of tests, and reporting to national authorities.